**On Surface Structure and Friction Regulation in Reptilian Limbless Locomotion**

H. A. Abdel-Aal

*Arts et Métier ParisTech, Rue Saint Dominique BP 508,*
*51006 Chalons-en-Champagne, France*
*Hisham.abdel-aal@ensam.eu*

**Abstract**

One way of controlling friction and associated energy losses is to engineer a deterministic structural pattern on the surface of the rubbing parts (i.e., texture engineering). Custom texturing enhances the quality of lubrication, reduces friction, and allows the use of lubricants of lower viscosity. To date, a standardized procedure to generate deterministic texture constructs is virtually non-existent. Many engineers, therefore, study natural species to explore surface construction and to probe the role surface topography assumes in friction control. Snakes offer rich examples of surfaces where topological features allow the optimization and control of frictional behavior. In this paper, we investigate the frictional behavior of a constrictor type reptile, Python regius. The study employed a specially designed tribo-acoustic probe capable of measuring the coefficient of friction and detecting the acoustical behavior of the skin in vivo. The results confirm the anisotropy of the frictional response of snakeskin. The coefficient of friction depends on the direction of sliding: the value in forward motion is lower than that in the converse direction. Detailed analysis of the surface metrological feature reveal that tuning frictional response in snakes originates from the hierarchical nature of surface topology combined to the profile asymmetry of surface micro-features, and the variation of the curvature of the contacting scales at different body regions. Such a combination affords the reptile the ability to optimize the frictional response.

**Nomenclature**

| | |
|---|---|
| $A_{pl}$ | cross sectional area that the counter face material established upon indenting the skin |
| $A_{real}$ | real area of contact between the contacting region of the reptile and the substrate |
| $F_f$ | Friction force |
| $F_s$ | shear component of friction force |
| $F_{pl}$ | ploughing component of friction force |
| H | Hardness |
| R | Radius of curvature |
| Ra | Mean arithmetic value of roughness ($\mu$m) |
| $R_{ku}$ | Profile Kurtosis parameter |
| Rq | Root mean square average of the roughness profile ordinates ($\mu$m) |
| $R_{sk}$ | Profile skewness parameter |
| $R_T$ | Radius of curvature in transverse direction |

***Directions***

| | |
|---|---|
| AE-PE | Anterior Posterior |
| RL-LL | Lateral Axis |
| LR | Lateral right hand side |
| LL | Lateral left hand side |
| LF | Lateral forward, |

| LB | Lateral backward. |
| SB | Straight backward |
| SF | Straight forward |

**Acronyms**

| COF | Coefficients of friction |
| DFE | Differential Friction Effect |
| FTAR | Fibril Tip Asymmetry Ratio, |
| LBH | Leading Body Half |
| MTS | Mid Trunk Section |
| RFA | Ratio of frictional anisotropy, |
| TBH | Trailing Body Half |
| WLI | White light Interferograms |

**Greek symbols**

| $\Theta_L$ | Fibril-tip leading edge apex slant |
| $\Theta_T$ | Fibril-Tip trailing edge apex slant |
| $\tau$ | Shear strength of the skin |
| $\mu_B$ | Coefficient of friction in backward motion |
| $\mu_F$ | Coefficient of friction in forward motion |
| $\mu_{T.H}$ | Coefficient of friction for the trailing half of the skin |
| $\mu_{L.H}$ | Coefficient of friction for leading half of the skin |

**Introduction**

The ultimate goal of surface customization for rubbing applications is to improve lubrication, reduce friction losses, and to minimize (or eliminate if possible) mass loss due to wear and friction-induced structural degradation in general. The design considerations for a surface depend on the particular tribological situation. In a lubricated surface, for example, it is desirable to alter the topography of the surface so that a full hydrodynamic regime is established within a short distance from the entrance of the lubricant to the rubbing interface (*Ferguson and Kirkpatrick 2001*). This leads to establishing complete separation of the rubbing surfaces early on in rubbing. Controlled adhesion may be a goal of surface structuring. Additional tribological design targets may be to establish anisotropic friction for motion control (e.g. for reduction of locomotion costs in rescue robots), or to control the wettability of a surface for enhanced lubricity or self-cleaning purposes (Thor et al., 2011).

A structured surface for enhanced tribo-performance should posses several advanced features. One principal feature is the ability to *tune* the frictional response upon rubbing. That is, the surface should be able to adapt its' frictional profile in response to sensed changes in sliding conditions (e.g. changes in texture of the mating surface, variation in contact pressure, etc.). The tuning requirement may stem from geometry of the surface topographical building blocks, their distribution and placement within the surface, presence of embedded sensory, or a combination of all these factors. One of the difficulties in engineering such a surface, despite the availability of several enabling technologies, is the current limited understanding of the interaction between deterministic surface textures and frictional response. This, in turn, is due to the relatively recent history of deterministic surface texturing in human engineering.

While the technical world lacks diverse examples of functional-self adapting tribo-surfaces, our surroundings contain an abundance of examples of hierarchically structured naturally occurring surfaces capable of delivering super functionality. These may provide inspiration for surface designers. The richness and diversity of the examples provided by natural surfaces are worthy of study to extract viable solutions for surface design problems encountered in the technical world. This is particularly feasible since the natural world obeys the same physical laws that govern the behaviour of engineering systems. As such, any extracted design rules should, in principle, be valid across both realms: the natural and the technological.

An order of species that manifest an interesting interaction between micro-structural surface features and frictional requirements for locomotion is that of snakes. Snakes belong to the serpents order within the Squamate Reptiles clade. Squamata (scaled reptiles) is a large order of reptiles of relatively recent origin. The order is distinguishable through the scales that are born on the skin of members of the order. Squamata comprises two large clades: Iguania and Scleroglossa. The later comprises 6,000 known species, 3100 of which are "lizards," and the remaining 2,900 species as "snakes" (Vitt et al., 2003). Snakes contain diverse examples where surface structuring, and modifications through submicron and nano-scale features, achieve frictional regulation manifested in: reduction of adhesion (Arzt., et al 2003) abrasion resistance (*Rechenberg, 2003*), and frictional anisotropy (*Hazel et al, 1999*). They are found almost everywhere on earth. Their diverse habitat presents a broad range of tribological environments.

Diversity in habitat requires adaptable features capable of efficient performance within the particular environment. Thus, a snake species particular to the desert, for example, would entail distinct features tailored to function within an abrasive sliding environment (Klein, et. al., 2010). The same would apply to a snake that roams a tropical forest where essential functional requirements differ from those dominant in a desert environment (Jayne and Herrmann, 2011). Function specialization requires analogous specialization in the composition, shape, geometry and mechanical properties of the skin. However, since the chemical compositional elements of reptilian skin are almost invariant within the particular species the study of functional specialization within a given species becomes more intriguing. This is because; invariance of chemical composition implies that functional adaptation takes place through adaptation of form, geometry and metrology of the skin building blocks. Such implications provide a venue to scour the customized surface features within the particular species to extract surface design lessons suitable for the technical world. Many, therefore, studied appearance and structure of skin in snakes Squamata (*Hazel et al, 1999, Vitt et al 2003, Chang et al, 2009, Alibardi and Thompson, 2002, Ruibal 1968, Chiasson et al, 1989, Jayne, 1988, Scherge and Gorb 2001, Rivera et al, 2005*). Furthermore, attracted by legless locomotion, others studied the tribological performance of snakes (*Berthe et al, 2009, Saito et al, 2000, Shafei and Alpas, 2008, 2009*). The results emphasized the role that diverse ornamentation actively contributes in the dynamic control of friction and regulation of locomotory energy consumption (*Shafei and Alpas, 2008, 2009, Abdel-aal et al, 2010, Abdel-aal and El Mansori, 2011, Gray 1946*).

In previous work Abdel-aal and co-workers (*Abdel-aal, et al, 2011, 2012*), reported the dynamic friction coefficient for the skin of a Python regius. The results confirmed the anisotropy of the friction of the reptile. The COF in forward motion (i.e., with the grain of the skin (caudal direction)) was less than that measured in backward motion (against the grain of the skin (cranial direction)). A similar trend emerged from measurements obtained in diagonal motion in both the forward and the backward directions while not reflected in measurements pertaining to lateral motion. The data suggested that such a friction differential effect stems from the geometry of the surface. In particular, the asymmetric profile of the individual micro-fibrils present on the ventral scales correlated to the anisotropy of the COF. Moreover, the metrological parameters of the surface (both macro and micro-scaled) showed a non-uniform distribution along the Anterior-Posterior axis of the reptile.

The variation in the metrological and geometrical parameters of the surface, in theory, affects the mechanics of contact between the ventral surface of the reptile and the substrate. Consequently, different locations on the body of the reptile will show varying frictional profiles. Additionally, due to the stocky build of the reptile, the distribution of body mass per-unit-length along the AE-PE axis of the body is non-uniform. Now, if the skin of the snake has a constant COF, then the non-uniform mass distribution will affect the frictional tractions and the friction-induced losses accommodated through the skin. That is, the irregular mass distribution will induce an analogous, locally variable, friction force distribution along the AE-PE axis. Such an irregularity should affect the structural integrity of the skin and on the energy consumed by the reptile to generate and maintain motion. However, observations in nature and in experimental work indicate that there is a distribution to the frictional tractions along the body of snakes (Berthe, et. al., 2009). In addition, the irregular distribution of frictional forces does not compromise the structural integrity of the skin. This is partly due to the ply-like skin structure (Klein, et. al., 2010). Interestingly moreover, the energetic cost of legless locomotion is found to be equivalent to that of running by limbed animals of similar size (Walton et., al., 1990). This implies that the COF of the skin varies locally and that an analogous distribution of the metrological parameters of the skin compensates the variation in the distribution of the frictional forces. In other words, the hierarchy of the textural features of the skin act as a control mechanism that "fine-tunes" the frictional response of the skin through modifying the contact between the reptile and substrate. In this sense, the skin of a snake not only would accommodate tractions (through its mechanical response) but also would actively control friction through texturing (micro-scale fibril elements). This is in contrast to recent explanations (Goldman and Hu, 2010; Hu et. al., 2009) that the overlapping arrangement of the ventral scales is the origin of frictional control and locomotion. This hypothesis, if validated, should contribute to linking the various textural elements (observed on snakeskins) to their tribological function. This, in turn, will help correlating the various shapes distributions, arrangements, of the micro-fibrils often observed on ventral scales of all snakes (Schmidt and Gorb, 2012) to their tribological environments and frictional response of the particular species. Such a correlation should enhance our understanding of the interaction between surface texturing and friction control in particular environments, thereby advancing our knowledge of surface engineering. To date, however, a study that examines such a hypothesis is non-existent despite its direct relation to many tribological problems, of fundamental nature, that relate to intrinsic control of friction and surface engineering.

The goal of this study is to compare local frictional behaviour of the skin to textural make up. Therefore, we investigate the validity of our hypothesis concerning the relation of surface texturing to frictional control in snakes. Namely, we attempt to answer the question of whether the COF for snakeskin is a property of the skin (whence a constant as implied in classical tribology) or rather it is a consequence of skin composition and particular texturing of the surface. Further, we investigate the correlation between surface geometry and the local variation in the COF along the body of the reptile. In principle, the present study is an extension of earlier studies by the author and co-workers. However, the current work comprises some fundamental differences that distinct the findings. Our earlier work (*Abdel-aal, et al, 2012*) stemmed from the premise that the COF of the skin is material property and therefore is a constant. Such an assumption implied that the geometry of the skin does not contribute to any functional adaptation. In addition, no attempt was made previously to link the frictional behaviour to the metrological features of the surface. Consequently, extrapolation of the findings to deduce design rules for technological surfaces was rather difficult. In the current work, however, the major assumption is that the skin of the snakes contributes to local adaptation through variation in micro-geometry. A consequence of such an assumption is that observed frictional behaviour of the snake, from a tribological point of view, is no longer a mere function of morphological traits (i.e., muscular activity). Rather, morphology and surface micro-design features form an integrated system of optimized, and adaptable, tribological function. Moreover, linking the frictional response to geometry should facilitate the deduction of design rules for technological surfaces especially that the description of surface topography is based on technological standards. This should facilitate the transfer of design ideas from the biological domain to the technological domain.

To simplify the presentation, without losing generality, we compare the metrological characteristics and the frictional behaviour for two locations on the skin of the reptile. The first location is representative of the leading half of the reptile, whereas, the second represents the trailing half. For brevity, we focus on presenting the metrological parameters that directly affect the frictional behaviour of the skin as implied from our preliminary study (*Abdel-aal et al., 2012, 2011*).

The manuscript comprises two parts. The first provides a comparison between the surface structure, geometrical and textural metrology on both sides of the skin. The second part, meanwhile, presents data pertaining to the frictional behaviour of each half of the skin and a comparison of the general trends emerging for each half. In addition, part two of the manuscript provides a correlation between the metrological parameters of the surface and the frictional behaviour determined in this work.

## 2. Anatomy of Snake skin

The skin of a reptile comprises two basic strata: the "dermis" and the "epidermis". The dermis is deeper than the epidermis. It is composed mainly of connective tissue. The epidermis contains an abundance supply of blood vessels and nerves. However, it does not have blood supply of its own. This renders the living cells, contained within this layer, depending in their nourishment on diffusion from capillaries in the dermis layer.

The epidermal layer in a snake entails seven layers. These are organized in plies of cells with tight packing. The epidermal layer (sub-layers included) encases the body of the retile to form an outer shield. Figure one details the seven layers present within the epidermis. Described from the inside of the skin, the first layer is the "stratum germinativum". This is the deepest layer of the skin. It is lined with rapidly dividing cells. Six additional sub-layers, again from the inside of the skin, follow. Together the six sub-layers form a so-called "epidermal generation" (old and new skin layers). Thus, stacked above the stratum germinativum, there exists: the clear layer and the lacunar layer. The lacunar layer matures in the old skin layer as the new skin is growing beneath. Following there is the α–layer, the mesos layer and the β-layer. The mesos layer is similar to the human stratum corneum (*Fraser and Macrae, 1973*) and contains several layers of flat and extremely thin cells surrounded by intercellular lipids (*Lillywhite and Maderson, 1982*). These three layers consist of cells that become keratinized with the production of two types of keratin: α (hair-like) and β (feather like). Keratinisation continuously transforms these cells into a hard protective layer. Finally, there is the "oberhautchen" layer, which forms the toughest outer most layer of keratinized dead skin cells.

The oberhautchen layer contains the fine surface structure known as the micro-ornamentation (*Meyers et al, 2008*). Before the molt, a new layer of epidermis forms under the currently existing one; the two layers are zipped together by a spinulae structure (Alibardi and Toni, 2007). During the molt, the reptile sheds the outer (older) layer of the epidermis. The principle constituents of snakeskin are keratin fibres (Toni et al; 2007). Keratinized-cells constitute the outer part of the skin. The process of keratinisation consists in synthesizing keratins that will potentially form the keratin fibres. The keratinisation brings an increase of keratin production from the cells that start to begin platter before dying (Ripamonti et al., 2009).

Two types of keratins form the epidermis: the α-keratin (which in a snake is acid or neutral) and the β keratin (which in a snake is basic). The β-layer consists mainly of β-keratin; this type of keratin is not present in other layers of the skin. Alpha α-keratin constitutes most of the epidermal layers and it contributes to the mechanical properties of skin cells (Maderson 1985).

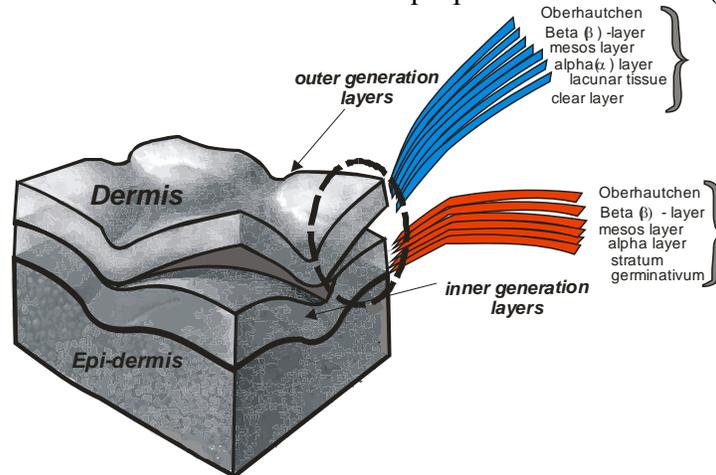

*Figure 1: General structure of the epidermis of a squamate reptile. The figure depicts the "outer" generation layer which is the layer about to be shed; and the "inner" generation layer which is the new replacement skin layer.*

The oberhautchen consists mainly of β-keratin. The presence of the two different types of keratinaceous protein α and β distinguishes the reptilian epidermis from its' mammalian counterpart (Fraser and Macrae, 1973). The shed epidermis of snakeskin consists of four layers: the outermost Oberhautchen, the β-layer (mainly protein), the mesos layer (lipid-rich), and the inner α-layer (mainly protein). The oberhautchen consists of a particular type of β-cells that play a major role in the shedding process. Together, this layer and the β-layer, both containing β-keratins, are considered as a unique β-layer in the mature epidermis. The oberhautchen layer is the outermost ply within the epidermal layers. It contains the micro-textural ornamentation. It is also the layer which in direct contact with the surroundings. That is it is the most active layer of the skin in the sense that it simultaneously accommodates contact and frictional effects.

## 3. Materials and Methods

### 3.1 Skin treatment

All observations reported herein pertain to shed skin obtained from five male Ball pythons (Python regius). All the received shed skin was initially soaked in distilled water kept at room temperature for two hours to unfold. Following soaking, the skin was dried using compressed air and stored in sealed plastic bags. Note that the exuvium surface geometry of shed epidermis does not differ from that of a live animal (*Klein et al, 2010, Klein and Gorb, 2012*). Therefore, using shed skin to characterize of the skin contribution to the frictional response should not affect the quality of the results.

## 3.2 surface texture metrology

Evaluation of surface texture metrology utilised a white light interferometer (WYKO 3300 3D automated optical profiler system). Analysis of all resulting White light Interferograms, WLI, to extract the surface parameters used two software packages: Vision ®v. 3.6 and Mountains® v 6.0. To determine the metrological features of the skin, we identified three major regions on the hyde of each of the studied snakes, these are shown in figure 2. The first region is located at the mid-section of the reptile. It is about 20 cm long and is the stockiest portion of the trunk (contained in the dashed rectangle in the figure). This was termed the Mid Trunk Section (MTS).

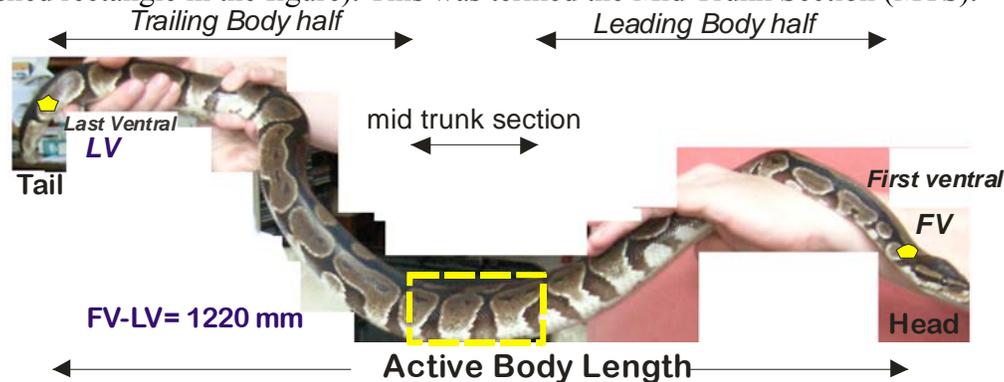

*Figure 2 positions chosen on the snake shed skin for metrological characterization.*

The remaining portion of the active length of the skin was then divided in two parts roughly equal in length (L= 47 cm). The portion of the skin extending from the first ventral scale (point FV in figure 3) to the right hand side boundary of the MTS was labeled as the Leading Body Half (LBH). The portion of the skin extending from the left hand side of the MTS to the Last Ventral scale (point LV in figure 2) was labeled the Trailing Body Half (TBH).

For each of the skin halves, we recorded fifty WLIs at randomly selected points within the particular half of the skin. These were further analyzed to extract the textural metrological parameters. In this work, we did not examine the MTS since we considered its geometry an anomaly with respect to the rest of the body. However, work currently in progress is comparing the makeup and friction behavior of this section to the rest of the body.

## 3.3 Friction measurements

All friction measurements utilized a tribo-acoustic probe, which is described elsewhere All measurements utilized a patented bio-tribometer (Zahouani et al., 2009). The device includes a tribo-acoustic probe that is sensitive to the range of friction forces and the acoustic emission generated during skin friction. It is also capable of measuring normal and tangential loads and of detecting sound emission due to sliding. The probe comprises a thin nitrocellulose spherical membrane, 40 mm in diameter, with a thickness of 1 mm. The probe material has a Young's modulus of 1 GPa. The roughness of the probe, $R_a$, is 4 μm and the mean value between peak to valley, $R_z$, is 31 μm.

In all frictional tests, the skin was stationary and the tribo-probe was moving at an average speed of 40 mm/s using a normal force of 0.4 (±0.05) N. The skin used in measurements consisted of 150 mm long patches taken from four locations on the ventral side of the shed skin. Skin samples did not receive any chemical or physical treatment beyond the water-assisted unfolding procedure described in the previous section.

The skin used in measurements consisted of 150 mm long patches taken from four locations on the ventral side of the shed skin, two from the leading half of the skin and two from the trailing half of the skin. Skin samples did not receive any chemical or physical treatment beyond water unfolding.

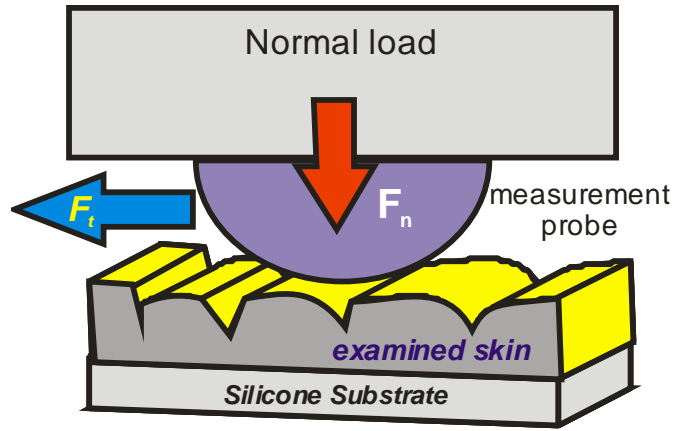

*Figure 3: sample setup of the skin sample and the tribo-acoustic probe used in measurements .*

To mimic the effect of the body of the snake on the skin, before starting an experiment, the particular skin patch was placed on a rectangular elastic pad of dimensions length L= 200 mm ,width W=100 mm and thickness of approximately 4 mm (figure 3). The pad is made of silicone rubber (Silflo®™, Flexico Developments Ltd., Potters Bar, UK). Table 2 provides a summary of the pad material properties.

Measurement of the friction forces proceeded along the two major body axes: the anterior-posterior axis (AE-PE) and the lateral axis (LL-RL) (see figure 4-a). In addition, we performed measurements along the diagonal directions shown in figure 4-a. For each direction, measurements were taken in the forward and backward directions. Figure 4, depicts the sense of forward and backward in relation to the motion of the reptile first in a global sense (figure 4-b), and second as it applies locally n the ventral scale (figure 4-c). To facilitate the description of the results we provide table three, which describes the measurement directions on each of the examined skin halves in vector form.

Table 2: Summary of geometric dimensions and mechanical properties of elastic pads used to cushion skin in experiments.

| Geometry | Rectangle | |
|---|---|---|
| | Length (mm) x Width (mm) | 200 x 100 |
| Material | Silicone Rubber (Silflo ®, Flexico Developments Potters Bar, UK) | |
| Mechanical Properties | | |
| | Young's Modulus (E) MPa | 2 @20 C |
| | Poissons Ratio | 0.3 |
| | Stiffness (K) | 300 N/m |

Table 3: Summary of vectors representing direction of friction measurements

| Measurement Direction | Vector Designation | |
|---|---|---|
| **Principal Directions** | *Leading Half* | *Trailing half* |
| Caudal | C.O →AE | PE→C.O |
| Cranial | C.O →AE | PE→C.O |

| Dextral | LL → RL | RL → LL |
| Sinistral | RL → LL | RL → LL |
| **Diagonal Directions** | | |
| Dextro-Caudal | C.O → AE-LL | PE-RL→ C.O |
| Sinistro-Cranial | AE-LL → C.O | C.O → PE-RL |
| Sinistro-Caudal | C.O → AE-RL | PE-RL→ C.O |
| Dextro-Cranial | AE-RL→ C.O | C.O → PE-LL |

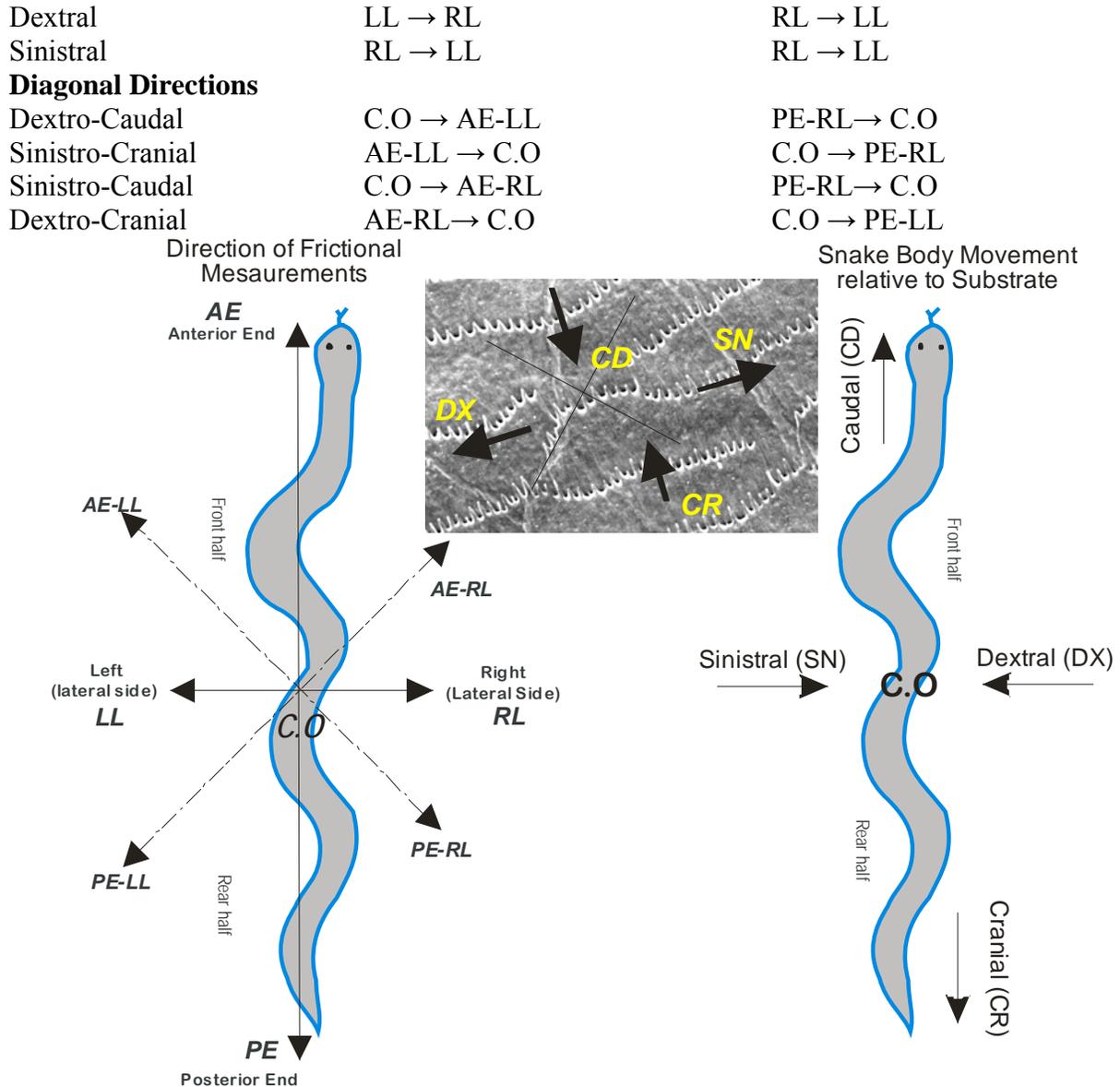

*Figure 4 Description of the axes-used to define directions of frictional measurements on the skin of the reptile (axes are defined in table 3in vector form)*

## 4. Metrological Characterization

The results of our preliminary study (*Abdel-aal et al., 2012*) identified parameters pertaining to surface asperity height, asperity distribution and form as primary metrological quantities. In this work, therefore, we will limit the presentation of the metrological aspects of the shed skin to those parameters.

### 4.1 Small Scale Metrology

As described earlier, in section 3, initial metrological characterization of the skin took place by generating White Light Interferograms (WLI) of selected patches within the ventral scales of the shed skin. Figure 5 depicts two of such WLIs. The interferogram shown as figure (5-a) depicts the overall topography of a ventral skin patch located within the leading half of the skin, whereas that shown in figure 5-b, details a patch located within the trailing half of the skin. Processing each of the interferograms provided profile data along the directions used for frictional measurements. The remaining plots, within figure 5, depict the extracted roughness profiles in the following directions: AE-PE-Axis (figure 5-c, d), RL-LL-Axis (figure 5-e, f), AE-RL-PE-LL-Axis (figure 5-g, h), and AE-LL-PE-RL-Axis (figures 5-i, j). Note that the profiles presented in figures 5-c: 5-j) represent roughness along a line and not an area. The scale to the left of figures 5 c-j differs from that to the right of figures (5-a and 5-b). The numbers on the former represent the heights and depths of the surface protrusions with respect to a reference line (whence the positive and negative values). The numbers to the right of figures 5-a and 5-b represent absolute height of surface points (i.e., height is referred to the lowest point on the surface). The maximum peaks and valleys of the surface roughness, irrespective of the direction of profile extraction, does not exceed two microns (i.e.,$-2 \ \mu m \leq h \leq 2\mu m$). Additionally, on average, the differences in heights between the leading and the trailing halves of the skin are not pronounced. This observation, however, is rather deceptive as the statistical roughness height parameters, $R_a$ (average roughness) and $R_q$ (root mean square height parameter) shows some variation both with respect to the direction of profile extraction and with respect to the location of the examined patch on the ventral side (leading or trailing).

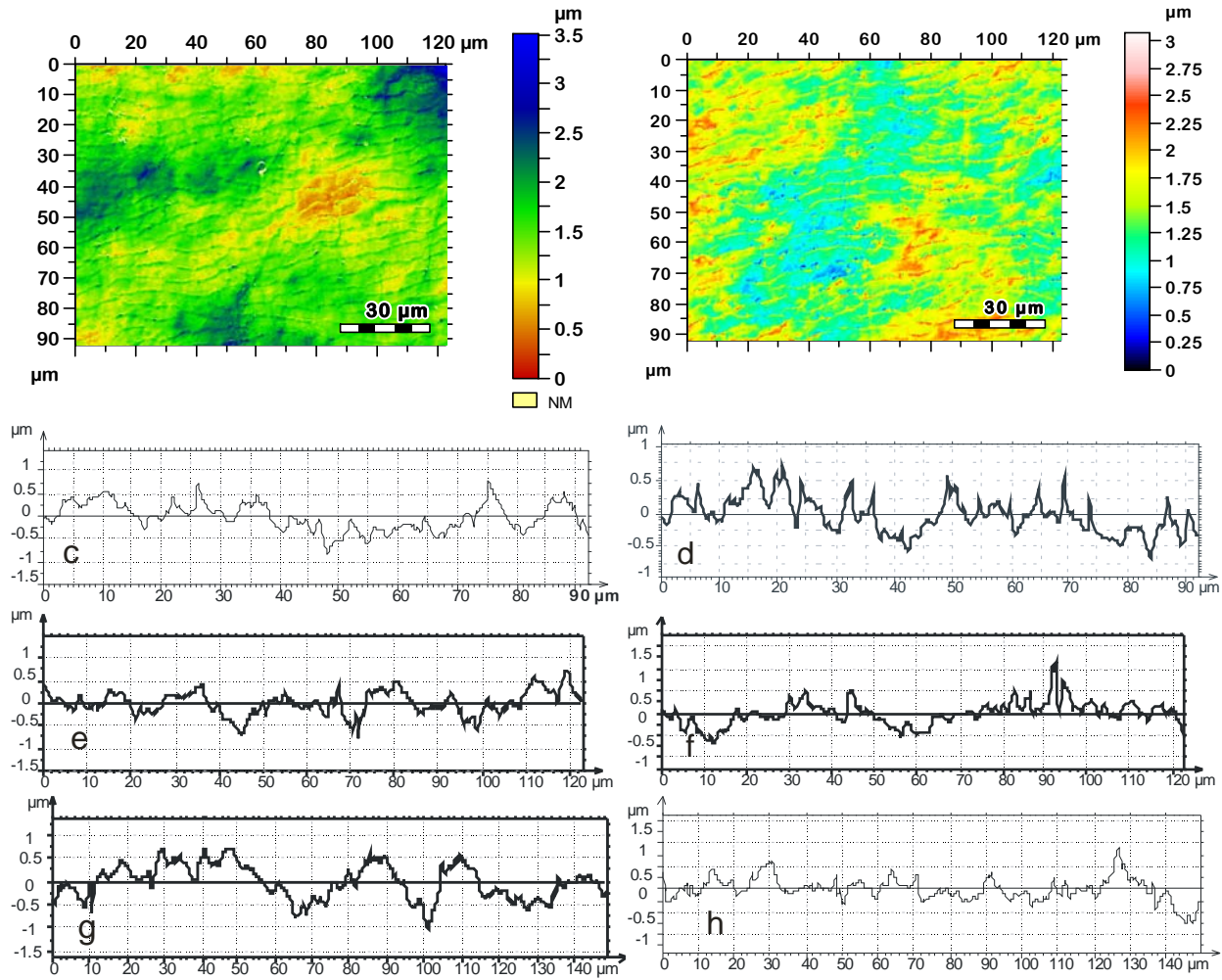

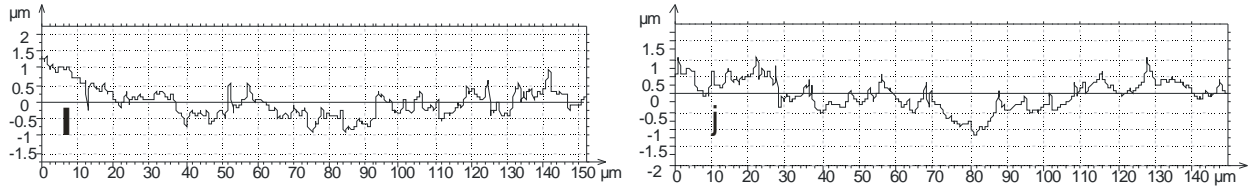

*Figure 5 Roughness profiles along the various axes of frictional mesaurements. The left hand side (figures 5-a, c, g, and i) depicts profiles representative of the leading half of the skin and the right hand side (figures 5-b, d, h, and j) depicts profiles representative of the trailing half of the skin.*

This variation is illustrated in figure 6 (a-d). The figure depicts bar plots of two roughness parameters extracted for each of the profiles shown in figure 5 (c through j). The plotted parameters are the mean arithemetic roughness parameter $R_a$ (figure 6- a and 6-b); and the root mean square average roughness parameter $R_q$ (figures 6-c and 6-d). The extracted parameters are plotted for the leading and the trailing half of the skin for ease of comparison. Examination of the data reveal both parameters, $R_a$ and $R_q$, have similar magnitude ranking. Namely, the highest values of these parameters pertain to the profiles located along the pricncipal axes (the Anterior poterior, AE-PE, and the lateral, LL-RL, axes). Values for profiles located along the diagonal axes are smaller than those along the principal axes. The values of the roughness parameters vary by location on the body of the reptile. For example, values of the roughness parameter Ra on the leading half of the skin are, in general, smaller than their counterparts on the trailing half of the skin (compare the values in figure 6-a to those in figure 6-b). Values of the mean square roughness Rq reflect a trend similar to that of Ra (compare values in figure 6-c to values in figure 6-d).

### 4.2 Shape of fibril tips

The microstructure of the ventral scales constitutes waves of micron-sized fibrils (Abdel-Aal et al., 2011). The shape of the tips of individual fibrils influences the frictional behavior of the skin. Hazel and co-workers (Hazel et al., 1999) suggested that the spherically asymmetric shape of the tips is the origin of the anisotropic frictional behavior they observed in their investigation of the skin of a Boa constrictor. This suggestion highlights the importance of characterizing the shape of the fibrils and the relation of that shape to frictional behavior. In this work, we use two metrics to characterize the shape of the fibril tips. The first is the extraction of the projection of the topography of a single fibril row in all directions of interest from WLI. The second is to map the profile kurtosis parameter $R_{ku}$ in all directions of interest.

### *4.2.1 Projection of Fibrils*

Figure 7 (a-d) shows the extracted profiles of a single fibril row. The orientation of all figures is inversed with respect to the natural position of the fibrils on the ventral scales and the position of the ventral side of the reptile during motion. Each profile shown in figure 7 is a plot of the projection of the fibril-tip in the respective direction. The apex of the profile is the point (or arc) that will contact the substrate when the reptile moves. Thus, in each of the figures, the top of the plot represents the relative position of the plane containing the contacting terrain with respect to the fibril plane. Based on this orientation the edges of a fibril are designated as "*leading*" or "*trailing*", and the motion is designated as "*forward*" and "*backwards*".

Figure 8 shows that fibril tips have an asymmetric profile. Moreover, a common theme to all examined fibril rows is asymmetry of slopes. The slope of the trailing edge of a fibril is steeper than the slope of the leading edge. The degree of profile asymmetry, however, is not constant. Rather, it varies with the direction of examining the profile. The variation of asymmetry seems to play an important role in determining the frictional profile of the reptile. To this effect, it is necessary to develop a quantitative measure to characterize the extent of fibril tip asymmetry. For such purpose, we define two angles $\Theta_L$ and $\Theta_T$. These angles denote the leading and the trailing edge apex angles of the fibril tips. Table three, presents a summary of the trailing and leading apex angles in all examined directions. The Values are in degrees. The last column within the table gives the ratio of the leading to the trailing apex angles $\Theta_L/\Theta_T$. The values of the angles confirm the observation that the leading angle is greater than the trailing angle in all directions.

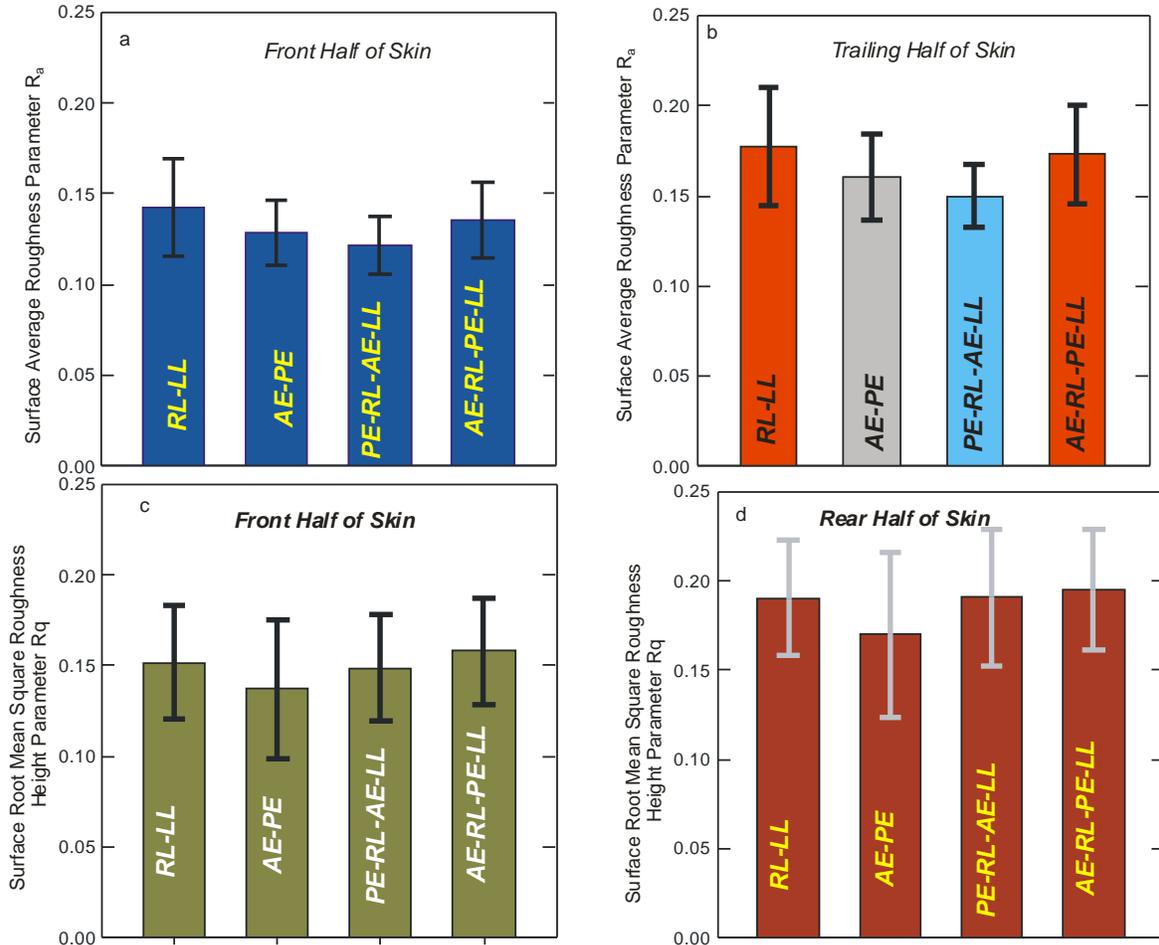

*Figure 6 Distribution of the average and root mean square roughness along the axes of measurements for the leading and the trailing halves of the skin. Figures 6-a and 6-b depict distribution of Ra representative of the leading and the trailing halves of the skin respectively. Figures 6-c and 6-d depict distributions of Rq representative of the leading and the trailing halves of the skin respectively (error bars are ± SD, values are significantly different One way ANOVA, p< 0.001)*

### 4.2.2 Profile kurtosis

The kurtosis is a measure of the "peakedness" or "roundness" of the distribution of the asperity heights (the fibril tips in this work). It measures the number of surface peak measurements that significantly vary from the mean of the heights. High kurtosis values ($R_{ku} > 3$) indicate a surface with a very wide distribution of surface heights, with many high peaks and low valleys (a so-called spiky surface). A low value ($R_{ku} < 3$) meanwhile implies a surface that is relatively flat, with the majority of the asperity heights close to the mean (a so-called bumpy surface). For a Gaussian (perfectly random) surface, the kurtosis parameter $R_{ku}$ is equal to three (Whitehouse, 1994). The value of the kurtosis of a surface affects the friction force developed during sliding. In particular, the contact loading is directly proportional to the value of the kurtosis (Tayebi and Polycarpou, 2004). When the kurtosis of a surface is high, more asperities establish contact with the counter-face body. This increases the real area of contact and, in turn, increases the frictional force. At very low values of the kurtosis, adhesion dominates especially in relatively smooth surfaces (Liu et al., 1998). Figure 8, presents a plot of the kurtosis parameter $R_{ku}$ in the directions examined in figure 5. Figure (8-a) presents the kurtosis values for the profiles located within the leading half of the skin whereas, figure 8-b depicts the $R_{ku}$ values for the trailing half of the skin.

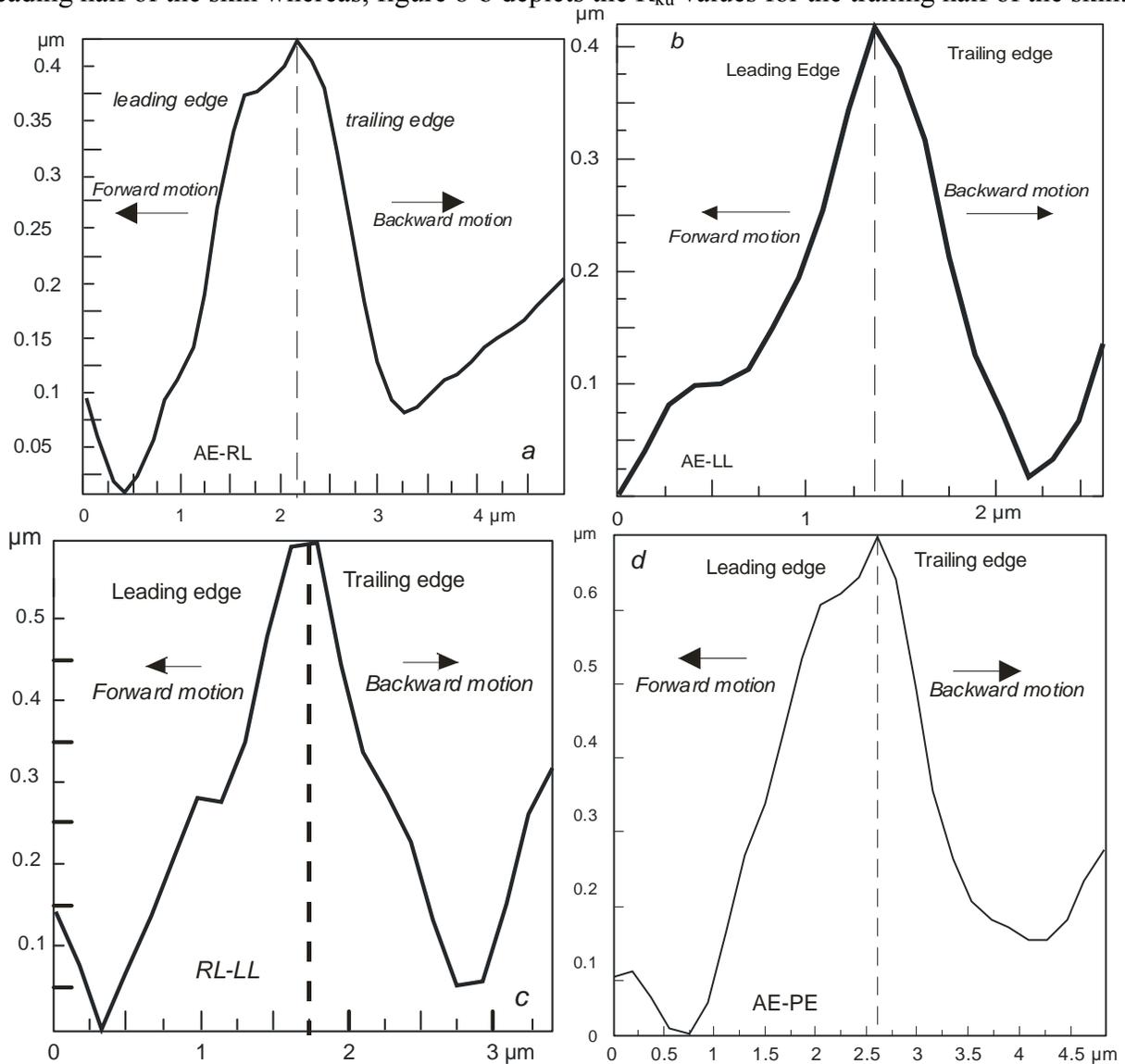

*Figure 7 Fibril tip profile along the different axes of measurements. Note that "forward" and "backward" directions of motion depend on the projection of the fibril raw with respect to the reptile axis (refer to table 3 for detailed description of directions of motion)*

Table 3 Definition and summary of fibril tip angles.

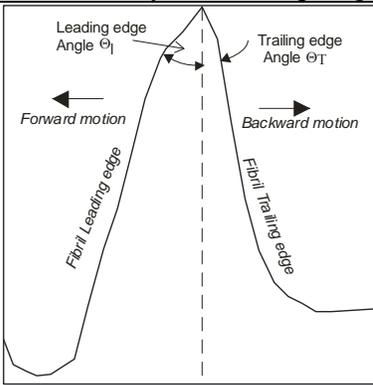

| Direction | $\Theta_L$ (deg) | $\Theta_T$ (deg) | $\Theta_L / \Theta_T$ |
|-----------|------------------|------------------|------------------------|
| AE-RL | 21.65 | 14.8 | 1.46 |
| AE-LL | 16.4 | 17.2 | 0.95 |
| RL-LL | 14.4 | 11.6 | 1.24 |
| AE-PE | 21.25 | 13.3 | 1.6 |

The plots imply that values of the $R_{ku}$ parameter depend on the orientation of the particular profile. Moreover, the magnitude ranking of the values does not display a consistent order. For example, within the leading half of the skin (figure 8-a), the smallest kurtosis value pertains to the profile along te AE-PE-Axis. This, however, is not the case within the trailing half of the skin, figure 8-b, where the smallest value pertains to one of the diagonal profiles (along the dextro-cranial direction AR-PL). Similarly, the largest kurtosis value within the leading half is that of the diagonal profile AL-PR (sinistro-caudal axis), whereas, the largest value within the trailing half is that of the lateral profile (LR-LL).

Values of the kurtosis within the leading half of the skin are, in general, less than three (the cut-off value for complete Gaussian height distribution) except for the sinistro-cranial profile. In contrast, within the trailing half of the skin, the kurtosis values are greater than three ($R_{ku} > 3$) except for one diagonal profile for which the value is very close to three. The data of figure 8 imply that fibril-tip heights within the trailing half have a random distribution. For the leading half, moreover, the kurtosis values fall within the interval ($2.25 \leq R_{ku} \leq 3.25$). This implies that the leading half of the skin is generally more flat than trailing half. As such, other factors being the same, friction of the leading half would entail higher adhesion contribution than the trailing half.

## 4.3 Form and curvature

Surface topography profiles presented in figure 5 are a superposition of two components: the basic form of the skin surface (the so-called deterministic component of roughness), and the rugosity (the so-called stochastic component of roughness). Separation of the form component provides information about the periodicity of the basic surface constituents and about the symmetry of the surface structural elements. This information, in turn, contributes toward understanding the kinematics of the surface during locomotion. One way of looking at form information, is considering contact of mating surfaces. Form establishes the global geometry of the contact area (ellipse, circle, etc.,). Rugosity, however, modifies this shape (e.g, by inducing deviation from basic shapes or causing the discontinuity of the contact spot). Figure 10 depicts the general form of an individual ventral scale located at the upper bound of the trailing half of the skin (scale was replicated using silicone rubber). Figure 9-a depicts a confocal microscopy scan of the ventral scale, whereas figure 9-b represents WLI form extraction of the shed skin.

Figure 9-a, shows that the ventral scale is convex toward the substrate. The convexity is not uniform throughout the surface of the scale. Rather, as illustrated in figure 9-b, the radius of curvature in the dorsal-ventral plane is smaller than that in the anterior-posterior plane ($R_T < R$). Consequently, the curvature of the scale surface is greater in the dorsal-ventral plane than in the anterior-posterior plane ($1/R_T > 1/R$). The asymmetric convexity of the scales affects the contact mechanics of the reptile.

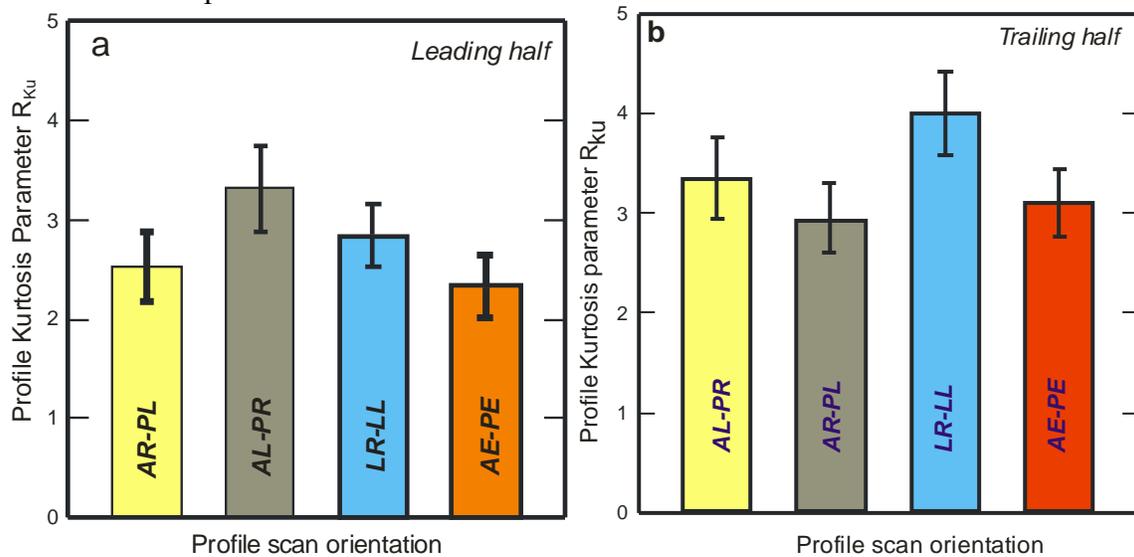

*Figure 8   Comparison between the values of the kurtosis parameter $R_{ku}$ in several directions. Figure 8-a depicts kurtosis distribution representative of the leading half of the skin; figure 8-b depicts kurtosis distribution representative of the trailing half of the skin (error bars are $\pm$ SD, values are significantly different One way ANOVA, p< 0.001)*

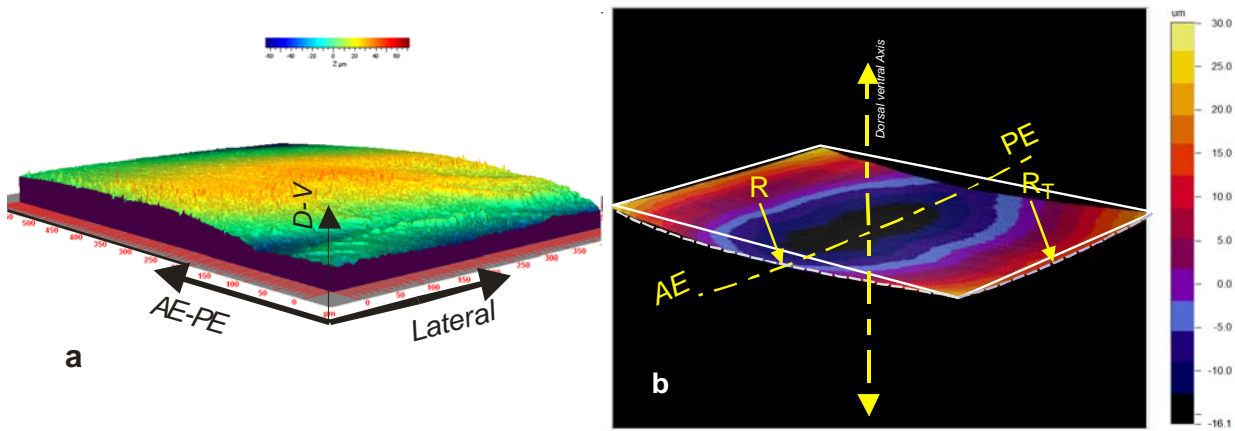

*Figure 9  General form of a complete ventral scale a- complete scanning using confocal microscopy ( ventral scale was replicated using Silicone rubber)and b- Band Fourier filtering of white light interferogram*

The displacement of the reptile mainly takes place along the AE-PE axis; the increased curvature will lead to minimizing the contact area between the body of the reptile and the substrate (in comparison to the contact area in case of a scale with a flat surface). Amonotons' law of friction implies that the friction force developed at the interface is proportional to the real area of contact. As such, if the area of real contact between the snake and the substrate is minimized the frictional forces will also be minimized. Reduction of the friction force reduces the energy requirement for locomotion (again compared to the case of a ventral scale with a non-curved surface).

To investigate the local curvature of the ventral scales along the body, we examined six additional ventral scales located at various zones within the skin. Figure 10 depicts the selected locations on the ventral side along with the extracted form and curvature information. The spots labeled (a-c) are located on the leading half of the skin, whereas, the spots labeled (d-f) are located on the trailing half of the skin.

Each of the spots depicted in the figure represent a square patch on the skin of approximate dimensions (1 mm x 1 mm). To extract curvature data, first the particular ventral scales were replicated using silicone rubber (see table 2 for mechanical properties of the replication material).

The use of Silicone to replicate the scales was favored to direct examination of shedskin due to the nature of the data to be extracted. In form extraction, the data required is several orders of magnitude larger in length than the roughness data (several hundred microns for the former and several microns for the later). In addition, the examined surface, the entire scale, has relatively large geometrical dimensions (area exceeds 150 mm$^2$). Replication of skin using silicone rubber provides the structural stability needed for sample examination by WLI. It is to be noted that skin replication using silicone rubber is frequently used in dermatological, metrology, and tribology studies of skin. The material and the procedure used in the current study is capable of capturing the finest metrological details of the skin specimens (*Asserin, et al, 2000, Jacobi, et al, 2004, Rosen et al, 2005, Forslind, 1999*).

Further to replication, we recorded WLIs for the zones of interest. Finally, curvature was evaluated by first filtering the WLI for form data then performing FFT on the filtered image.

The figure shows that the local (small-scale) curvature, similar to the global (form curvature) depends on the location on the body. The radii of curvature in the AE-PE and the lateral axes are not constant, nor equal. Towards the tail (within the trailing half), the curvature is almost one-dimensional (especially close to the tail where the body is curved in the lateral direction only). For the spots located on the leading half of the skin, curvature is two-dimensional however; the radii of curvature are not equal.

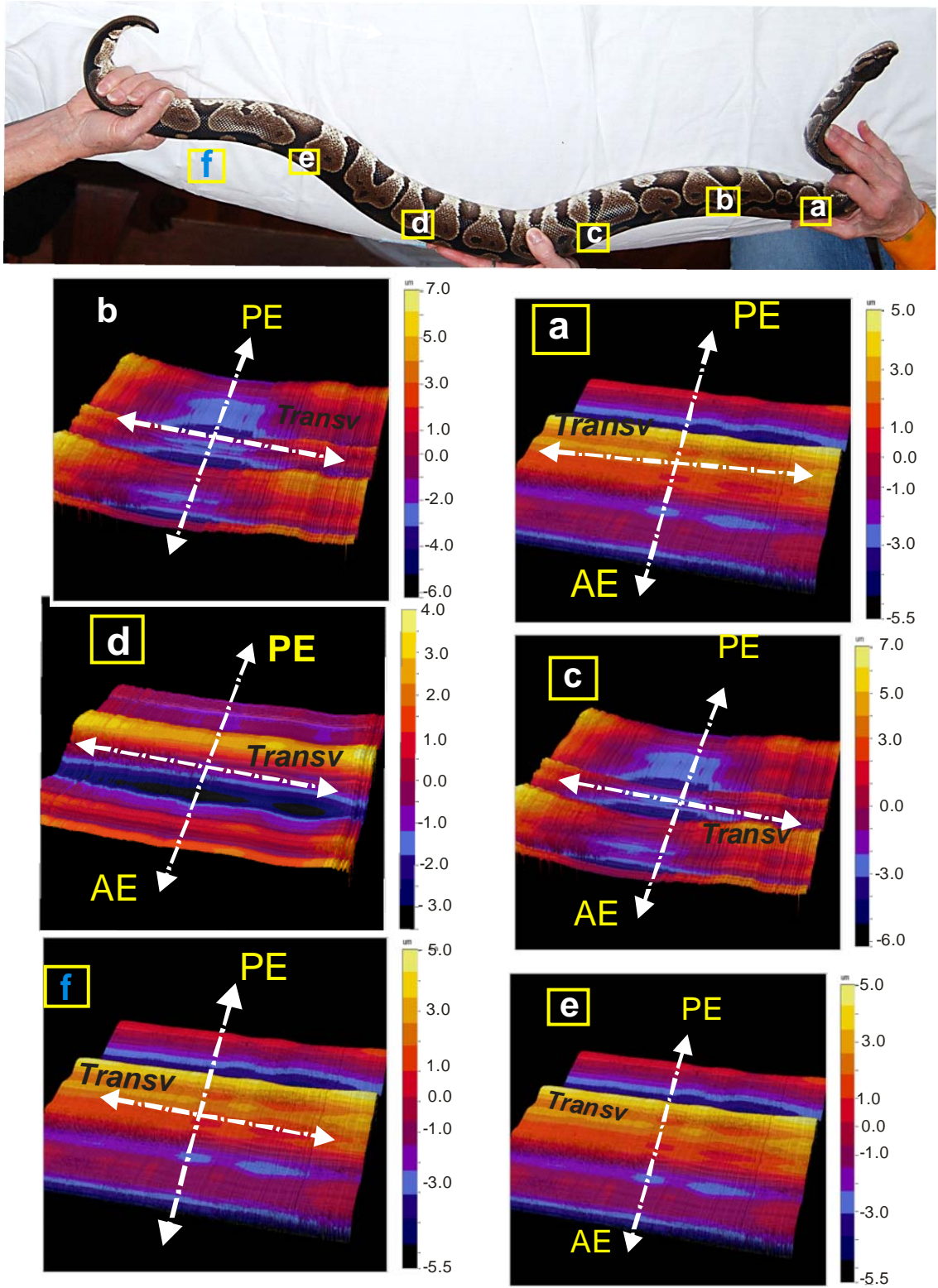

*Figure 10 variation of the curvature of the ventral scales for six 1 mm by 1mm skin spots located at different region of the skin. Note the change in curvature within the trailing portion of the skin.*

## 5. Friction Measurements

This section presents the results of friction measurements obtained on each of the examined skin halves. Conditions of measurements were summarized earlier in section 3.4. For each skin patch used in the experiments, frictional measurements were performed in eight directions. One hundred and fifty measurements were taken for each direction (ten measurements on each of three different positions on each half, on skin from each of five different individuals). All data were statistically analyzed using SigmaPlot® version 11.0. Kruskal-Wallis one way ANOVAs followed by Tukey tests with a significance level of p,0.001 were performed.

Figure 11, presents a summary plot of the COF obtained in all measurement directions for both halves. The left hand side of the figure depicts a plot of data obtained for the leading half of the skin. Data for the trailing half, meanwhile, are plotted on the right hand side of the figure.

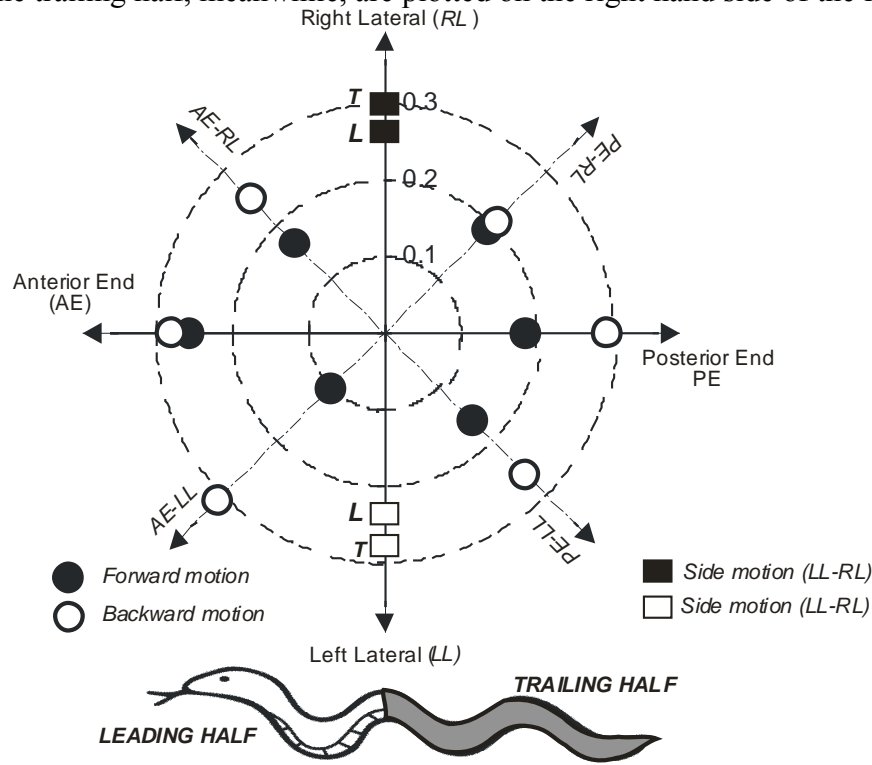

Fig. 11 *Distribution of COF values for each of the measurement directions for leading and trailing halves of the skin.*

Closed symbols denote COF values for forward motion (refer to figure 4), whereas open symbols denote measurements in backward motion. Square symbols dente values of the COF obtained in lateral motion (along the Sinistral-Dextral direction). Each point within the plot represents the statistical average of ten measurements. The data indicate that friction of the skin is anisotropic. Such anisropy is manifested in the COF for forward motion (in all directions) being less than that in the converse (backward) direction. Frictional anisotropy appears to be the dominant trend regardless of the orientation of measurements and regardless of which half of the skin is examined. Note that, for all measurement directions, and on both halves of the skin, $\mu_B$ is greater than $\mu_F$. The COF along the lateral axis (square symbols), however, contrasts the general trend of frictional anisotropy. The COF in the sinistral direction is roughly equal to the COF measured in the dextral direction.

To investigate the frictional anisotropy further we plotted the ratio of the backward to the forward COF ($\mu_B/\mu_F$) against the direction of measurement. The plot, shown here as figure 12, follows the same order of presentation used in figure 11. As such, data for the leading half of the skin are plotted on the left hand side, whereas, data for the trailing half are plotted on the right hand side.

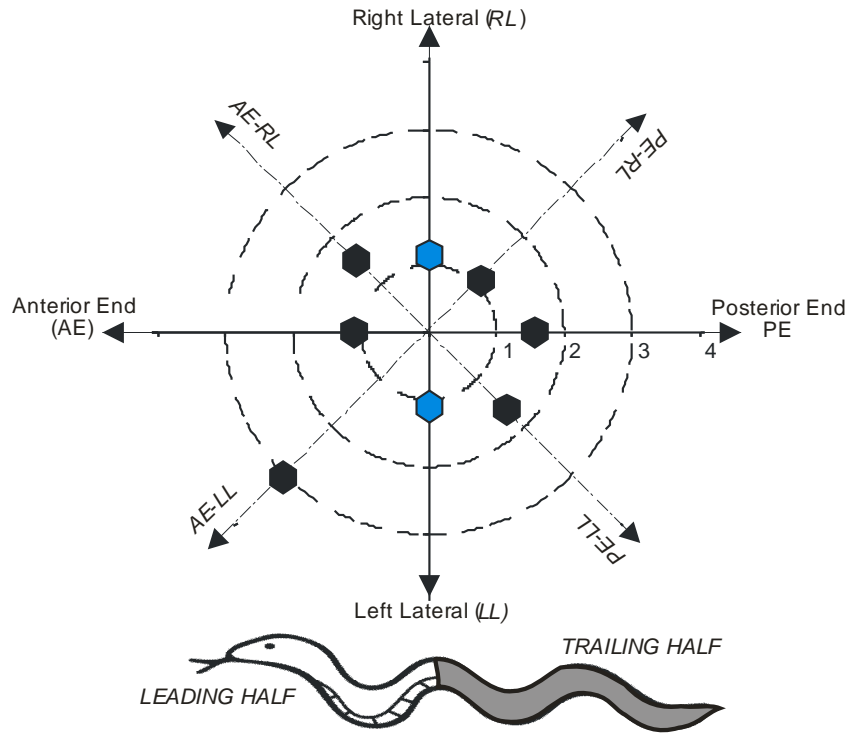

*Figure 12 Mapping of the frictional anisotropy ($\mu_B/\mu_F$) on the skin*

The plot shows that the anisotropy of the COF is not uniform. For example, consider the frictional anisotropy along the anterior-posterior axis (AE-PE). Along that axis the anisotropy within the leading half is less than that within the trailing half ($\mu_B/\mu_F \approx 1.2$ for the former while $\mu_B/\mu_F \approx 1.6$ for the later). In the lateral direction, however, the frictional anisotropy is almost invariable. The COF in the sinistral direction roughly equals the COF in the dextral direction.

Note that the absence of frictional anisotropy in the lateral direction does not imply that the individual values of the COF, within both halves of the skin, are equal. Rather, the absence of anisotropy pertains to the invariance of the ratio $\mu_B/\mu_F$. For diagonal motion, the highest anisotropy pertains to measurements taken along the dextro-caudal direction (and its converse).

It is of interest to compare the individual values of the COF on both of the examined skin halves. Figure 13 presents such a comparison. The figure depicts the ratio of the COF within the trailing half to its corresponding value of the leading half (i.e., the ratio $\mu_{T.H}/\mu_{L.H}$). Open symbols within the figure represent the ratio $\mu_{T.H}/\mu_{L.H}$ for motion in the backward direction (along the particular axis). Closed symbols, meanwhile, denote the same ratio but for motion in forward along the particular axis. The plot implies that the COF in forward motion for the trailing half of the skin is higher than that within the leading half ($[\mu^F_{T.H}/\mu^F_{L.H}] > 1$). The inverse is noted for the COF in backward motion. For this set of measurements, the ratio is less than one ($[\mu^{B}_{T.H}/\mu^{B}_{L.H}] < 1$).

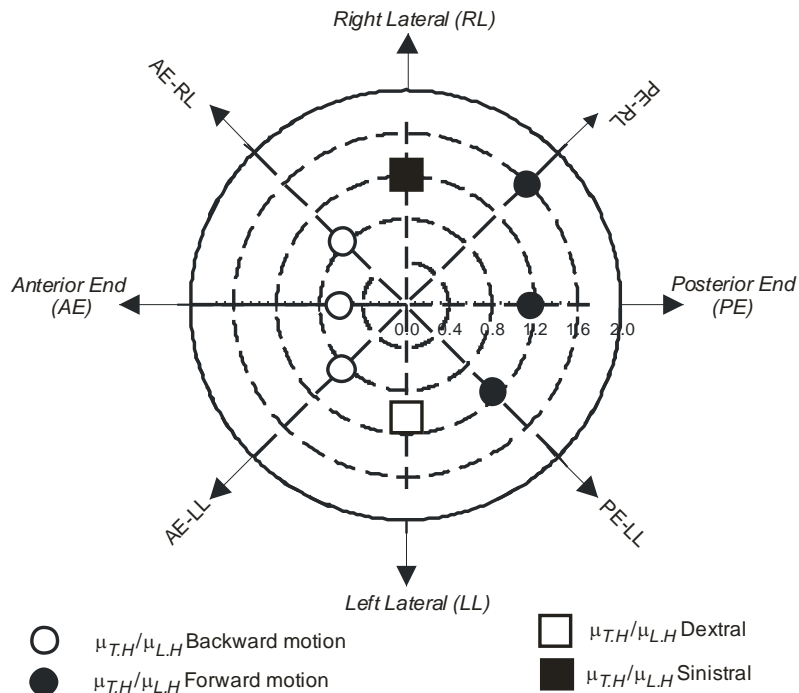

*Figure 13 Comparison of COF values within the trailing half to its corresponding values of the leading half ($\mu_{T.H}/\mu_{L.H}$)*

Measurements of the COF in lateral motion (i.e., along the LL-RL- axis) display an interesting behaviour with respect to friction resistance to motion. The COF in the sinistral direction within the trailing half of the skin is higher than that measured for the leading edge (i.e., $\mu_{T.H}/\mu_{L.H} \approx 1.2$, closed square symbols). This implies that when the reptile slides from the LL end toward the RL end of the skin ventral scales within the trailing half of the skin offer more resistance to motion than ventral scales located within the leading half. However, for motion in the dextral direction, the frictional resistance offered by each of the skin halves is almost equal (i.e., $\mu_{T.H}/\mu_{L.H} \approx 1$, open square symbols).

## 6. Discussion

### 6.1. Comparison to previous work

One of the principal findings of the current study is confirmation of the anisotropic behaviour of the COF. This anisotropy, not only existed in relation to the direction of motion, but also it was reflected in relation to position on the body (trailing half or leading half). Keeping in mind the method used to evaluate the COF in this work, and that the measurements entailed the use of shedskin and not a live animal, it is of interest to compare the current data to data reported in other studies. Table 4, presents a compilation of the COF obtained by several researchers along with data of the current work. Figure 14 (a and b) meanwhile depicts a graphical correlation of the ratio of frictional anisotropy, $\mu_B/\mu_F$, and the lateral COF $\mu_{lat}$, to relevant compiled data.

Table 4 Comparison between COF measurements obtained in the current study and values reported in open literature

| Substratum | $\mu_F$ | $\mu_B$ | $\mu_B/\mu_F$ | $\mu_{Lat.}$ | Species | Reference |
|---|---|---|---|---|---|---|
| Metal (dry) | 0.33 | 0.33 | 1 | | Grass snake *Natrix Tropidonotus Natrix* | Gray and Lisseman, 1950 |
| Sandpaper | 0.65 | 0.75 | 1.15 | | | |
| Fine | 0.65 | 0.88 | 1.35 | | | |
| Medium | 0.44 | 1.31 | 2.98 | | | |
| Rough | 0.61 | 1.32 | 2.16 | | | |
| Epoxy Resin Ra (μm) | | | | | Amazon Boa *C. hortulanos* | |
| 0.08 | 0.207 | 0.227 | 1.09 | 0.23 | | Berthe et al, 2009 |
| 0.25 | 0.218 | 0.222 | 1.02 | 0.24 | | |
| 0.42 | 0.203 | 0.204 | 1.05 | 0.22 | | |
| 1.11 | 0.17 | 0.165 | 0.97 | 0.19 | | |
| 2.26 | 0.153 | 0.153 | 1 | 0.19 | | |
| 2.75 | 0.161 | 0.162 | 1.01 | 0.19 | | |
| 12.67 | 0.144 | 0.157 | 1.09 | 0.21 | | |
| 13.94 | 0.151 | 0.181 | 1.19 | 0.22 | | |
| Styrofoam | | | | | Corn snake *(Pantherophis guttatus guttatus)* | Marvi and Hu (2012) |
| Conscious (static) | 0.51 | 0.88 | 1.72 | | | |
| Unconscious (static) | 0.3 | 0.35 | 1.17 | | | |
| Conscious (*dynamic*) | 0.49 | 0.79 | 1.61 | | | |
| Unconscious (*dynamic*) | 0.21 | 0.35 | 1.67 | | | |
| Glass Ball (1 mm diameter) Ra (0.006) μm | | | | | *King Snake L.G Californiae* | Benz et al (2012) |
| Soft Cushioned | 0.121 | 0.154 | 1.27 | 0.14 | | |
| Hard Cushioned nitrocellulose Ra =4 μm | 0.081 | 0.151 | 1.85 | 0.08 | | |
| Leading half | 0.256 | 0.281 | 1.09 | 0.26 | Python regius | current work |
| Trailing half | 0.185 | 0.293 | 1.58 | 0.27 | | |

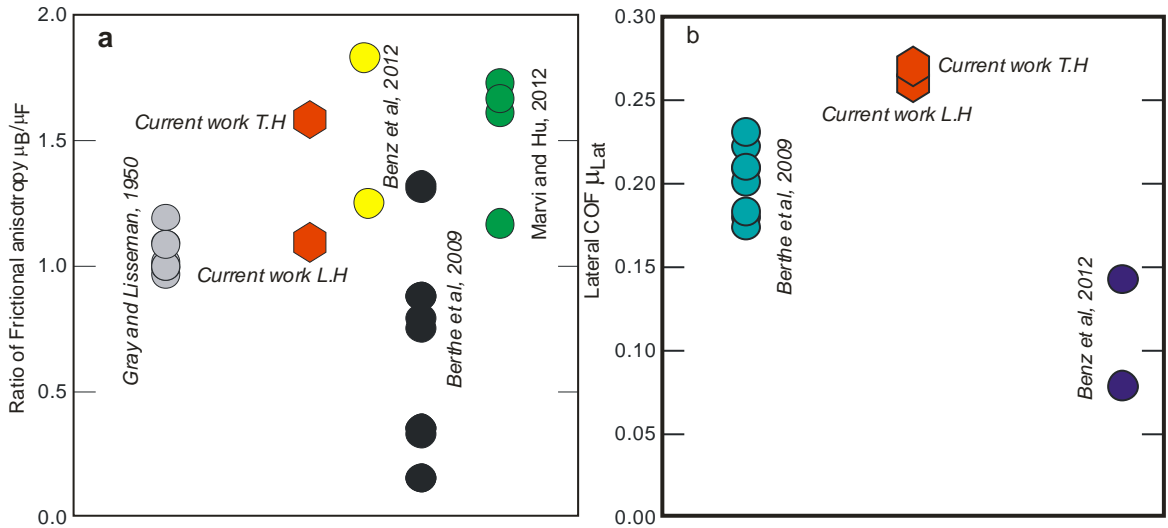

*Figure 14: Correlation of the ratio of frictional anisotropy $\mu_B/\mu_F$ and the laterl COF $\mu_{Lat}$ obtained in the current study with measurements reported in literature*

The data of table 4 reflect measurements performed on different species, and using different methods. The data also reflect frictional measurements for a variety of substratum materials. In general, the measurements of the current study agree with the compiled data. In particular, the compiled data support the direction based anisotropic trend of the COF. The degree of anisotropy, however, varies according to the material of the mating substratum, and its roughness values (observe the data for epoxy resin for example). It is interesting also to note that despite the variation in the individual values of the COF for the particular species, the ratio of the anisotropy remains very close for the majority of cases (see figure 14-a). The lateral COF, however, displays a curious trend. The differences between species seem to be of little influence on the magnitude of this variable (see figure 14-b). One possible explanation may be that the species used by Berthe (*Berthe, et al, 2009*) is a Boa, which is a large snake similar to the Python regius used in the current work with respect to eight and length. Both species meanwhile move using rectilinear locomotion.

Despite the agreement between the data of the current study and data of other researchers a note of caution is due. The agreement of the data shoud be cautiously accepted becase of the differences in the type of friction investigated in the current study and that investigated for most of the studies. The current study measures dynamic friction whereas most of the data reported in table 4, and plotted in figures 14-a, b) pertain to static friction. In addition, the number of available measurements in literature and the corresponding species investigated is rather lmited. The data available report on friction of ten (or slightly more) snakes. Clearly this is a limited number taking into account the approximately 3000 types of snakes currently present on the planet. It is plausible, however, that the COF for snakes may be bounded between two low limits, even more plausible that all measurements would orbit a certain range of values. However, in light of the severe lack of data such an assumption may not be adopted. Finally, the diagonal COF are reported in this work for the irst time in literature, for this reason direct comparison with results from other researchers was not possible.

## 6.2. Correlation of frictional behaviour to topography

The distribution of roughness (asperities) and the shape of asperity tips influence the frictional response of any sliding surface. In the current work, the parameters that represent asperity-tip distribution and asperity geometry are the profile kurtosis, $R_{ku}$, and the Fibril Tip Asymmetry ratio, $\Theta_L/\Theta_T$ (defined in section 4.2.1). It is interesting, therefore, to correlate these variables to the COF measurements and to the observed frictional anisotropy observed along the various measurement orientations. Such a correlation is important on two counts. Firstly, it helps identifying the relationship between the geometry of the asperities and the anisotropic frictional behaviour of the surface. Secondly, the correlation offers an in-depth perspective of the intricacies of design of reptilian surfaces. This later aspect stands to enhance our fundamental understanding of frictional control through geometrical customization of surface texture. In other words, correlating the geometrical features of the surface micro-constituents to frictional measurements facilitates the deduction of design rules that possibly could apply to the conception of manufactured functional surfaces.

Figure 15 presents a plot of the ratio of frictional anisotropy, RFA, for both halves of the skin, versus the Fibril Tip Asymmetry Ratio, FTAR. The figure contains plots of three quantities along the vertical axis. The plots represent the RFA for the leading half of the skin (circular symbols), the trailing half of the skin (hexagonal symbols), and the average value of the surface (triangular symbols). Along with these three plots the figure depicts the linear least square fit of the mean (average) RFA with respect to the FTAR.

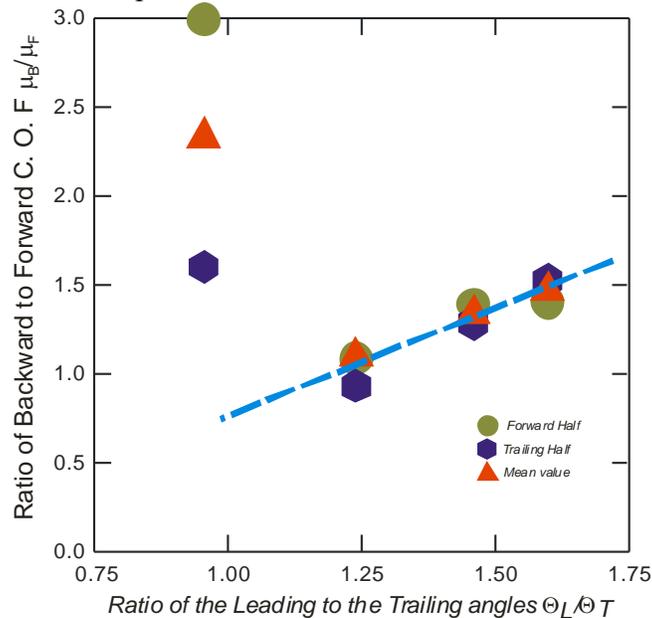

*Figure 15 Correlation of fibril tip asymmetry to frictional anisotropy. Note that linear fit was obtained without accounting for the outlier COF values ($\mu_B/\mu_F$=0.72 + 0.32 $\Theta_L/\Theta_T$, $R^2$ = 0.0569)*

In calculating the linear fit, we excluded the mean value obtained along the sinistro-cranial diagonal axis (refer to table 2). This is because such a value results from including the RFA in the C.O-AE-LL direction ($\mu_B/\mu_F \approx 3$, refer to figure 12), which is deemed as an outlier value with respect to the rest of the measurements.

The data imply a linear relationship between the RFA and the FTAR. The linear relationship between the two quantities indicates that frictional anisotropy of the skin is a consequence of the asymmetric geometry of the fibril-tips. Such a conclusion supports the earlier findings of Hazel et al (1999) who reported frictional anisotropy upon studying the friction of Boa Constrictor skin. Hazel and co-workers suggested that the presence of fibrils within the scales aids the reptile in conditioning its frictional response through a ratcheting effect. This ratcheting effect originates from the symmetry of fibril-tip geometry.

The asymmetric profile of the fibril tips causes the fibrils to have different slopes. The slope of the tips is gradual in the direction of forward motion and rather steep in the direction of backward motion. This asymmetrical tip-shape, as suggested by those authors, acts as a ratchet. This ratchet action results in a differential friction effect manifested in the resistance to motion of the reptile in the forward direction being less than that in the backward direction (whence the frictional anisotropy). However, Hazel and co-workers did not quantify the fibril structure in terms of metrological parameters, or fibril-tip exact geometry. This later quantity is reported here for the first time in tribology or biology literature.

The ratcheting scenario suggested by hazel and co-authors, offers a rudimentary explanation of the origin of the observed anisotropy. Other work by this author, currently in preparation, implies that frictional anisotropy is a manifestation of stick-slip sliding motion induced by the geometry of the fibril-tips. In this stick-slip process, significant energy dissipation takes place through friction-induced acoustical emission. Thus, due to the presence of the fibrils, roughness-induced sound emission takes place upon sliding. Presence of the fibrils, leads to the emission of almost uniform acoustical waves when the reptile moves forward. However, upon attempting to move backwards, motion is not continuous. Rather, due to asymmetry of the tips, stick slip motion takes place. This causes additional dissipation of locomotion energy through instantaneous damping of the emission in the stick phase of motion. In such a case, the energetic cost of backward locomotion for the reptile is prohibitive.

Figure 16 depicts the variation in friction with the kurtosis parameter $R_{ku}$. The figure comprises two plots. Figure 16-a, depicts the variation in the RFA with the profile kurtosis parameter, $R_{ku}$, whereas, 16-b depicts the variation in the COF.

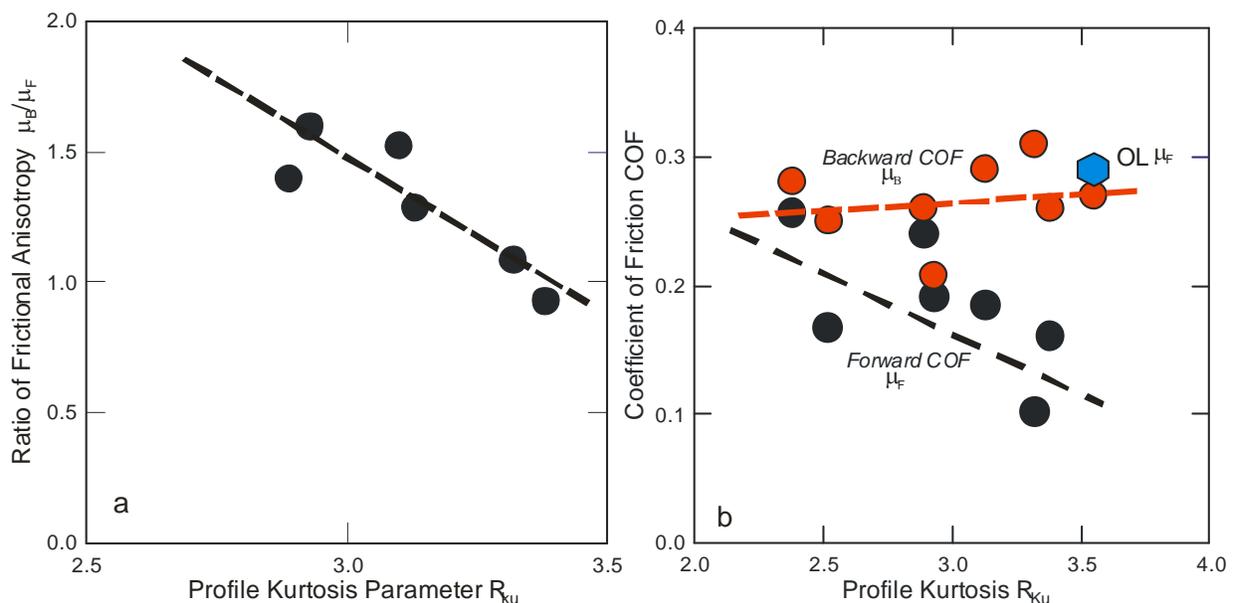

*Figure 16 Correlation of frictional anisotropy to profile kurtosis. Figure 15-a depicts the variation in the RFA with the profile kurtosis parameter, $R_{ku}$. ($\mu_B/\mu_F$=1.85-0.174 $R_{ku}$, $R^2$= 0.461) Figure 15-b depicts the variation in the COF measurements with the Kurtosis.( $\mu_B$=0.2157+0.017 $R_{ku}$,$R^2$=0.05294; $\mu_F$=0.252-0.0177 $R_{ku}$,$R^2$=0.0147)*

The dashed lines in figure 16-a represents a linear least square fit of the RFA data. Again similar to the approach adopted in figure 15, we considered that the RFA in the AE-LL direction is an outlier. As such, this particular data point was not included in obtaining the linear fit. The plot shows that the RFA is inversely proportional to the profile kurtosis. Recalling that the Kurtosis represents the spatial distribution of the asperities, a linear correlation implies that the fibril tips contribute to the anisotropy of friction. An interesting observation, however, is that the RFA is inversely proportional to the kurtosis. Thus, frictional anisotropy increases upon the decrease of the profile kurtosis parameter.

In manufactured surfaces, an increase in the kurtosis leads to an increase in the number of contacting asperities. Generally, an increase in the number of contacting asperities increases the area of contact. The increase in asperity numbers, in turn, increases the frictional force acting at the interface. Such an increase, however, should not be confused with the trend implied in figure 16-a. The decrease displayed by the COF, at higher kurtosis values, pertains to the RFA and not to individual values of the COF. However, to investigate whether the COF follows a trend similar to that of the RFA we plotted the obtained measurements against their respective $R_{ku}$ values. This plot is presented as figure 16-b. The figure presents the plots of two data sets. The first, denoted by red circles, pertains to the COF measurements, whereas the second, denoted by black circles, pertains to forward measurements. The dashed lines represent a linear least square fit of the variation in the particular data set with profile kurtosis. In consistence with the development of figure 16-a, regression lines in figure 16-b were calculated without including the outlier COF value (hexagonal symbol).

The data display contrasting trends. The backward COF measurements slightly increase with the increase of profile kurtosis. Such a trend is consistent with the general trend observed in manufactured (manmade) surfaces. For forward measurements, however, the COF decreases with the increase in profile kurtosis. Such behaviour contrasts COF observations in manufactured surfaces and points at possible influence of the profile skewness, $R_{sk}$. Such a point is currently a subject of further investigation. In made-made surfaces, the so-called lay of the surface is a function of the manufacturing process itself, and that causes, among other factors, the well-known frictional behaviour reported by Tayebi and Polycarpou. For natural surfaces, however, the origin of the contrasting behaviour is not quite clear. One possible reason may be the randomness of the surface lay in case of snakes. In fact preliminary data obtained by the author points at the random (or isotropy of the surface lay in natural surfaces), however in light of the lack of conclusive data such a reasoning may not be confirmed.

The effect of the skewness parameter on the COF contrasts the effect of the kurtosis (especially for lightly loaded rubbing surfaces such as snakeskin). Accordingly, for snakeskin the COF should be inversely proportional to the skewness. Frictional behaviour of heavily loaded rubbing surfaces, however, does not depend on the skewness and the COF is such case is almost constant (Tayebi and Polycarpou, 2004). To verify such an effect, we evaluated the skewness values for the roughness profiles used in this work. Figure 17 presents a plot of the COF measurements against profile skewness ($R_{sk}$). Again, the figure includes plots of the forward and backward COF values along with linear fits. The plots show that the COF is proportional to the skewness. This trend is common to forward, as well as, backward measurements. In light of the positive values of the skewness, $R_{sk} > 0$, the decrease in the COF points at the dominant role that profile tip geometry has on the friction behaviour of the ventral scales. To this end, we propose that the fibril structures, present within the ventral scales, have a dominant influence on the frictional behaviour, and tuning, of the reptile. Such a proposal is in contrast with recent views that the overlapping structure of the ventral scales in snakes is the origin of frictional control and friction anisotropy (Hu et al., 2009, Goldman and Hu, 2012).

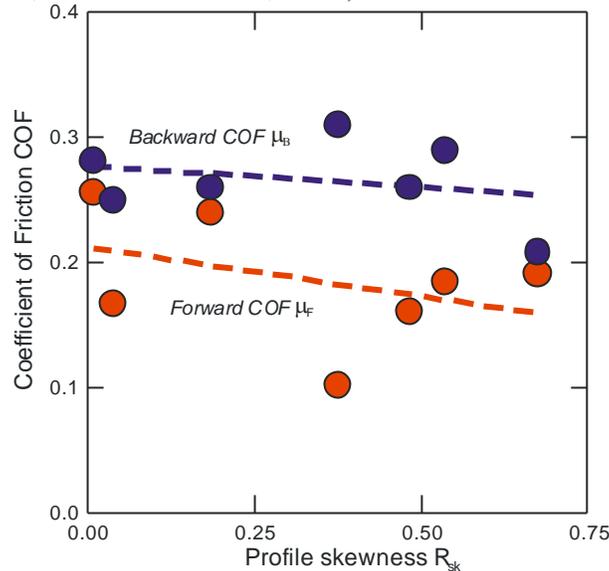

*Figure 17 Variation in COF measurements with skewness parameter Rsk ($\mu_B$=0.277-0.034 $R_{sk}$,$R^2$=0.0694, $\mu_F$=0.2117-0.078 $Rs_k$,$R^2$=0.1522)*

To illustrate the role of the fibrils in modifying the friction force we invoke Tabors analysis of the fiction force (Bowden and tabor, 1958).

The friction force that develops between the ventral side of the reptile and the sliding substrate consists of two components. The first is a shear component, $F_s$, whereas the second is a ploughing component, $F_{pl}$.

$$F_s=\tau A_{real} \tag{1-a}$$

Whereas, the ploughing component may be calculated from:

$$F_{pl}=H A_{pl} \tag{1-b}$$

Where $\tau$ is the shear strength of the skin (assumed to be softer than the substrate), $A_{real}$ is the real area of contact between the contacting region of the reptile and the substrate, H is the hardness of the softer material (the skin), and $A_{pl}$ is the cross sectional area that the counter face material establishes upon indenting the skin. As such, the total friction force, which is the sum of the two components, takes the form:

$$\begin{aligned} F_f &= F_s + F_{pl} \\ &= \tau A_{real} + H A_{pl} \end{aligned} \tag{2}$$

Snakes move under the influence of their own body weight and the forces exerted by their respective locomotory muscular groups. The contact pressures resulting in this case are low. Under such conditions the real area, $A_{real}$, and the indentation area, $A_{pl}$, may be considered equal. This leads to further simplification of the expression for the friction force to,

$$F_f=(\tau+H)A_{real} \tag{3}$$

Equation (3) implies that when the material properties, shear strength and hardness remain unchanged, only the real area, $A_{real}$, will affect the frictional force. This allows the comparison of the friction force developed in case of a textured surface (i.e., a ventral scale that contains fibrils), and that developed in case of a non-textured surface, sliding on the same substrate. Thus, we write,

$$\frac{F_f^F}{F_f^{NF}} = \frac{A_{real}^F}{A_{real}^{NF}} \qquad (4)$$

The superscripts "F" and "NF" denote the presents of fibrils on the ventral scale and the absence of fibrils respectively. For identical contact pressures in both contact cases, the force ratio takes the form,

$$\frac{F_f^F}{F_f^{NF}} = \frac{\mu^F}{\mu^{NF}} \qquad (5)$$

Comparing equations (4) and (5) yields,

$$\frac{\mu^F}{\mu^{NF}} = \frac{A_{real}^F}{A_{real}^{NF}} \qquad (6)$$

Equation (6) implies that area of contact for a ventral scale containing fibrils will be lower than that established in the case of a ventral scale with no fibrils present. This leads to lower COF for the former compared to that for the later. Equation (6) highlights the role of fibril distribution along the Anterior-Posterior axis of the reptile in controlling friction. This non-uniform distribution, (*Abdel-Aal and El Mansori, 2011*), should affect the variation of the COF along the body. This in turn affects the energy consumption of the reptile in locomotion.

The motion of a snake is a delicate balance between the propulsive forces generated by the muscles and the friction tractions due to contact with the substratum. In some cases, the snake makes use of friction to generate thrust. However, in general the COF needs to be minimized (especially in rectilinear locomotion) since friction opposes motion. As such, a mechanism to control the frictional traction should exist in the snake. This mechanism is provided by the texture of the surface (i.e., the micron sized fibril structures). The use of the fibril structures allows the snake to modify the real area of contact. In other words, the presence of the fibrils results in the segmentation of the real area of contact between the skin and the substratum. The area of contact in such a case will be the sum of the contact areas between those active fibrils and the substratum. In the absence of fibrils, the real area of contact will result from the contact between the form of the body and the substratum. Due to the difference in size between the form and the micro-roughness, the area of contact in the second case will be larger.

Adams and co-workers (*Adams et al., 1990*) observed that other natural materials that have keratin composition, such as camel hair and horsehair, exhibit similar frictional behaviour to that of snakeskin. Such materials manifest a so-called Differential Frictional Effect (DFE). Due to such effect, the frictional work required by a fibre to slide over another fibre varies according to the direction of sliding. The work required for a hair to slide in the direction of tip-to-root is greater than the work required to slide in the converse direction. That is, for such materials friction is anisotropic.

Adams and co authors attributed the origin of the DFE to the geometrical make-up of camel and horsehair. In particular, these authors reasoned that the origin of asymmetrical friction for these materials is the presence of cuticle structures within individual hairs. The cuticle structures, which are similar to the micro-fibrils on a snake ventral scale, possess asymmetric topographical features that "resemble the way in which tiles are laid on a roof (Adams et al., 1990). An important feature of the work of Adams was to link frictional asymmetry to the non-uniform spatial distribution of the asperity tips of the cuticle structures. This non-uniform distribution, as suggested by Adams, limits the mechanical engagements between the surface of the hair and the substrate; and it provides autonomy in the way asperities interact with each other. The relative freedom of movement of the contacts, Adams argue, produces the DFE. Such a view supports the finding in the current work that the frictional anisotropy noted for snakeskin has a geometric origin. Indeed, the linear correlation between the RFA and each of: the asymmetry of fibril tip profiles, kurtosis, and skewness strongly confirms the geometric origin of frictional anisotropy. This finding provides an opportunity for designing functional surfaces that yield a predetermined frictional response by imposing surface textures of optimised kurtosis and asperity tip asymmetry.

**Conclusions**

In this work, we presented an experimental study of the frictional characteristics of the ventral side of reptilian shed skin (Python regius). The study compared the COF, and related metrological characteristics, on two regions of the ventral side of the reptile. Results showed that the COF depends on the direction of motion. In the forward motion (Tail-to-Head), the COF was less than that measured in the opposite direction (Head-to-tail). A similar trend was observed for COF values obtained in diagonal motion, on both the leading and the trailing halves of the skin.

The degree of frictional anisotropy, ratio of backward to forward COF was found to vary according to location on the body. Frictional anisotropy, on the leading half of the skin was, in general, less than that measured on the trailing half.

The ratio of asymmetry of fibril tips (ratio of leading edge angle to trailing edge angle) correlated to frictional anisotropy. The resulting relationship indicated that frictional anisotropy is a linear function of the geometric asymmetry of the fibril tip. This finding confirms the geometric origin of frictional anisotropy often observed for snakeskin. Two relationships resulted from the correlation.

Frictional anisotropy was found to vary linearly with profile Kurtosis. This suggests that the geometry of the surface strongly contributes to the control of friction in reptilian limbless locomotion. The control mechanism stems from an ability to vary the asymmetry and roundness of the contacting surface. Such a finding provides an opportunity to engineer surfaces with multi-scale features, which use variation in kurtosis, variable curvature, and asymmetry to control friction.